\documentclass[runningheads]{templates/springer/llncs}
\usepackage[T1]{fontenc}
\usepackage{graphicx}

\newenvironment{ResearchQuestion}[2]
{
  \vspace{3pt}
  {\narrower \noindent \textit{\textbf{Research Question #1:} #2}\par}
  \vspace{3pt}
}
{}
\usepackage{dblfloatfix}
\usepackage{tikz}
\usetikzlibrary{positioning, arrows.meta, fit, backgrounds}
\usepackage{booktabs}
\usepackage{multirow}
\usepackage{amssymb} 

\begin{document}


\title{Understanding LLMs in Title-Abstract Screening:\\ From Disagreements to Recommendations}

\titlerunning{Understanding LLMs in Title-Abstract Screening}

\author{Mika Mäntylä\inst{1}\orcidID{0000-0002-2841-5879} \and
Patricia Matsubara\inst{2}\orcidID{0000-0001-9230-3620} \and
Katia Romero Felizardo\inst{3}\orcidID{0000-0001-9080-4165
} \and
Miikka Kuutila\inst{4}\orcidID{0000-0002-3695-7280
} \and
Marco Gerosa\inst{5}\orcidID{0000-0003-1399-7535}\and
Savio de Sousa Sampaio\inst{6}\orcidID{0009-0009-1962-8635}\and
Tayana Conte\inst{6}\orcidID{0000-0001-6436-3773} \and
Igor Steinmacher\inst{5}\orcidID{0000-0002-0612-5790}}

\authorrunning{M. Mäntylä et al.}

\institute{University of Helsinki, Finland\\
\email{mika.mantyla@helsinki.fi} \and
UFMS, Brazil\\
\email{patricia.gomes@ufms.br} \and
UTFPR -- Federal University of Technology - Paraná, Brazil\\
\email{katiascannavino@utfpr.edu.br} \and
LUT University, Finland\\
\email{miikka.kuutila@lut.fi} \and
Northern Arizona University, United States\\
\email{\{marco.gerosa, igor.steinmacher\}@nau.edu} \and
UFAM, Brazil\\
\email{savio.sousa.sampaio@gmail.com, tayana@icomp.ufam.edu.br}}









\maketitle
\begin{abstract}
Several studies have examined the use of large language models (LLMs) for title-abstract screening in systematic reviews (SRs), reporting mixed accuracy. However, questions of reliability remain largely unaddressed. In this study, we go beyond quantitative LLM-human agreement metrics and qualitatively investigate how and why LLMs fail. We also propose actionable recommendations. We analyzed disagreements between LLMs and researchers across six software engineering SRs and over 1,000 primary study papers. For each SR, papers were screened independently by human experts and LLMs in zero-shot mode, resulting in Kappa values ranging from 0.52 to 0.77. Qualitative analysis suggests that human-LLM disagreement results from recurring, identifiable causes, such as boundary ambiguity in key terms, keyword overemphasization, and incorrect topic inference. Based on these findings, we propose recommendations such as validating semantic understanding before deployment, running multiple LLMs, and focusing validation efforts on borderline cases. Future studies are needed to validate the impact of our recommendations, and community efforts are needed to develop normative guidelines on LLM usage in SRs.

\end{abstract}

\section{Introduction}
An important step in systematic reviews (SRs) is title-abstract screening, when titles and abstracts of candidate studies are evaluated against inclusion and exclusion criteria~\cite{kitchenham2004evidence}. 
This step is labor-intensive, requiring researchers to analyze hundreds or thousands of records for a single SR. Moreover, the task is cognitively demanding and highly repetitive, making it prone to errors that may lead to the omission of relevant evidence, threatening the SR validity. This concern is reflected in the attention given to this step in established guidelines~\cite{petersen2015guidelines,kitchenham2004evidence,garousi2019guidelines}.

Recent advances in artificial intelligence have motivated studies in software engineering concerning the use of LLMs to support title–abstract screening ~\cite{huotalaPromiseChallengesUsing2024a,felizardoChatGPTApplicationSystematic2024,huotalaSESREvalDatasetEvaluating2025,petersen2025road,thode2025exploring,felizardo2025difficulties}. Beyond software engineering, several reviews have also examined this topic~\cite{madeyski2025llm4screenlit,lieberum2025large,kim2025evaluating,sandner2025evaluating}. These studies often report promising results, showing performance comparable to humans and beyond traditional NLP approaches in zero-shot setups. At the same time, existing evidence highlights concerns regarding reliability and variability across reviews, with agreement levels between humans and LLMs ranging from moderate to substantial~\cite{huotalaSESREvalDatasetEvaluating2025}.

Prior work has mostly focused on aggregate accuracy metrics and agreement statistics. However, these indicators provide limited insight into the causes of disagreement. Researchers adopting LLM-assisted screening need to understand not only \textit{how often} models disagree with humans, but also \textit{where} and \textit{why} this happens. Without that, it is difficult to design effective human–LLM workflows or mitigate risks associated with missed or incorrectly included studies.

To address this gap, we conduct a qualitative cross-study analysis of disagreements between LLMs and human raters during title–abstract screening. We analyze six software engineering SRs in which humans and LLMs independently screened the same candidate papers. Rather than treating disagreements as isolated errors, we examine their recurring structure and underlying causes. Specifically, we investigate the following research questions:

\begin{ResearchQuestion}{1}{What disagreement patterns between LLMs and human raters recur across studies? \label{sec:rq1}}
\end{ResearchQuestion}

\begin{ResearchQuestion}{2}{What recommendations can be drawn from the identified patterns? \label{sec:rq2}}
\end{ResearchQuestion}


The main contributions of this work include (i) a taxonomy that characterizes where and what goes wrong in LLM-assisted screening and (ii) a set of recommendations to help researchers anticipate and mitigate these disagreements in practice. Our results inform the design of more reliable screening workflows and provide concrete recommendations for researchers seeking to integrate LLMs into SR pipelines. These findings contribute to a more informed and reliable use of LLMs in evidence synthesis.

\section{Method}
In this section, we describe the subject studies and our approach to analyzing disagreements between LLMs and humans in title-abstract screening. We used a qualitative cross-study design to examine how and why these disagreements occurred. Humans and LLMs independently screened the same candidate papers from a set of SRs, after which we extracted all cases where their decisions diverged. We then qualitatively coded and synthesized these cases through inductive analysis to identify recurring disagreement patterns.


\subsection{Subject Studies}
To investigate disagreement patterns across diverse contexts, we selected a set of subject studies (SRs) spanning different software engineering topics and research groups. 
For all subject studies, we followed the same general procedure: papers were retrieved from Scopus using the original search strategies, human reviewers performed title–abstract screening and reached consensus, and then multiple LLMs screened the same sets in zero-shot mode. The LLM screening was done via the AISysRev tool~\cite{AISysRev} that allows screening of hundreds of papers input via a CSV file. The tool creates a zero-shot prompt for each paper by combining the title and abstract with the SR's inclusion and exclusion criteria, instructing the model to return a binary include/exclude decision along with a brief rationale.
All studies used at least two LLMs: \textit{gemini-2.5-flash} and \textit{openai/gpt-4.1-mini}. A recent study \cite{huotalaSESREvalDatasetEvaluating2025} found openai/gpt-4.1-mini to be the best-performing, although the differences among the top models were small. Overall, these models represent a modest token price while offering good performance.  

Although our work is not quantitative, we do provide statistics from all of our selected studies in Table \ref{tab:studies}. For readers who have more interest in LLM screening performance as a statistical phenomenon, we refer to the paper by Huotala et al \cite{huotalaSESREvalDatasetEvaluating2025} that contains quantitative analysis of over 34k studies in software engineering, or to the study by Chan et al that quantitatively analyzed screening decisions of over 540k studies in the medical domain \cite{chan2025comprehensive}. Our work's novelty is in the quality analysis, something that is not seen in past quantitative works. Next, we present the context of each subject study and study-specific characteristics.

\begin{table}[h]
\centering
\caption{Study statistics}
\label{tab:studies}
\begin{tabular}{lllllll}
\hline
\textbf{Metric} & \textbf{TP} & \textbf{ESEE} & \textbf{WOSS} & \textbf{GenAIEdu} & \textbf{AI4T} & \textbf{AI4TInd} \\ \hline
Primary studies                   & 137           & 100             & 100             & 100                 & 162             & 445                \\
Human include       & 15           & 6             & 38             & 8                 & 31             & 33                \\
LLM include         & 32           & 7             & 35             & 23                 & 21             & 43 (228)                \\
Cohen's Kappa       & 0.52           & 0.75             & 0.56             & 0.71                 & 0.72             & 0.77 (0.133)                 \\ \hline
\end{tabular}
\end{table}

\textbf{Time Pressure (TP)} -- This study corresponds to a previous SR on Time Pressure in software engineering ~\cite{kuutila2020time}. We searched for papers that were published after the SR, thus, it made us as unbiased as we had not previously made screening decisions on these papers. The search yielded  137 new papers from Scopus. Two human raters screened all studies and resolved disagreements through discussion. LLM-based screening was then applied. 
Then, both humans independently documented their observations on the reasons for the LLMs' divergence, which informed our final taxonomy. Overall, 11\% of the papers were included in the title-abstract screening phase. Cohen's kappa between the human consensus and LLMs was 0.523, indicating moderate agreement. For this calculation, the LLM decision was determined by whether any LLM included the study.

\textbf{Expert Judgment in Software Estimation (ESEE)} -- This systematic mapping study investigated factors that affect expert judgment in software effort estimation~\cite{matsubara2022sextamt}. The researchers used a random sample of 100 papers retrieved using the original search string in Scopus. Two researchers independently screened the sample and produced a consensus set of six included and 94 excluded papers. Next, LLMs were used 
to analyze the same set of papers. A paper was considered included by the LLMs if at least one model selected it. This resulted in seven papers selected for inclusion by the LLMs, with a Cohen's Kappa of 0.753, indicating substantial agreement between humans and LLMs. 

\textbf{Women in OSS (WOSS)} - This study corresponds to a previously published SR of women's participation in open source software~\cite{trinkenreich2022women}. For the present analysis, 100 papers were retrieved using the original search string in Scopus. Two researchers screened the studies and reached consensus on a set of 38 included papers. Subsequently, LLMs 
analyzed the same set, with at least one model selecting 35 papers for inclusion. Agreement between human consensus and the LLMs yielded a Cohen's kappa of 0.558, indicating moderate agreement. 

\textbf{Generative AI in Education (GenAIEdu)}-–-This unpublished systematic mapping study investigates the usage of Generative AI in Software Engineering Education.
The study analyzed 57 primary studies published from 2023 to 2025. For the present selection process, we used a random sample of 100 papers retrieved from Scopus. Title–abstract screening was conducted by four researchers, resulting in a final set of 8 included papers. LLMs  
analyzed the same random sample and selected 23 papers for inclusion (considering papers selected by at least one LLM). The agreement between human reviewers and the LLMs, measured using Cohen’s kappa, was 0.709, indicating substantial agreement. 

We included two additional ongoing studies to complement the previous four. These studies processed a larger volume of papers, offering breadth to enrich the analysis. However, they were evaluated only by the first author and LLMs.  
    
\textbf{AITest2023 (AI4T)} is a 2023 subset of papers from an ongoing SR evaluating the use of \textit{AI in test generation}. The year 2023 was selected to narrow down the number of papers to be suitable for this study. 
The first author independently screened a subset of 162 papers from the year 2023 and scored all papers for inclusion and exclusion. Then, LLMs screened the same papers, after which the first author checked for conflicts, resulting in a final set of 31 included papers. The LLMs included 21 papers (considering papers selected by at least one LLM). The final agreement between human and LLM, with Cohen's kappa of 0.727, is in line with previous studies.



\textbf{AITestIndustry (AI4TInd)} is another subset of papers from an ongoing SR on \textit{AI use in test generation}. In this subset, papers that two humans had previously screened for inclusion were selected for further analysis (445 papers). The first author aimed to narrow this set by identifying all papers conducted in the "software industry" to present for an industry audience. Thus, this represented a rapid review case where the goal was to support software engineering practice, similar to the study by Pizard et al.~\cite{pizard2025using}. Three LLMs were used (\textit{gemini-2.5-flash}, \textit{openai/gpt-4.1-mini}, \textit{anthropic/claude-haiku-4.5}). A paper was considered for inclusion by the LLMs only if all three LLMs included it. This approach was motivated by the rapid review context, where the objective was to present only the most notable cases to a software industry audience. The LLMs included 43 papers, while the human judge included 33 papers. The Kappa agreement between LLM and human decisions was 0.77, indicating substantial agreement.

For full transparency, we must also note that a further motivation for the LLM inclusion rule change was the fact that \textit{gemini-2.5-flash} highly underperformed in this task. It interpreted nearly any type of industry relevance argument in the paper abstract as an indication that the study had been conducted in the "software industry." Had we adopted the approach of using two LLMs (\textit{gemini-2.5-flash}, \textit{openai/gpt-4.1-mini}) and included a paper if either LLM deemed it relevant, we would have faced an excessive number of papers for qualitative analysis. Using those rules, the LLMs would have included 228 papers, and the Kappa agreement between LLM and human decisions would have been 0.133.

For each subject study, we extracted all instances in which any LLM decision diverged from the human consensus decision. The human consensus was treated as the reference judgment because it resulted from independent screening followed by conflict resolution. Each disagreement instance included the paper metadata, the human decision, the LLM decision, and any rationale produced by the models when available.

\subsection{Analysis}
All cases of disagreement were first open-coded to capture the preliminary reason for divergence. One researcher then analyzed the full set of coded instances and accompanying notes. She derived themes and their descriptions using inductive thematic analysis \cite{braun2006using}. Then three other researchers reviewed the themes and descriptions, along with the codes and notes associated with them. Next, the researchers discussed codes and themes, as well as the qualitative distinctions between themes, to further refine them. Finally, a fifth researcher conducted an additional review, renaming them to follow the <where–what> format and merging or creating themes as needed, resulting in the final list of general themes.
We did not compute a formal inter-rater reliability coefficient for the qualitative coding, as we used open coding and followed a consensual validation approach: themes were iteratively refined through discussion among researchers until agreement was reached on the assignment of all cases. 

Throughout the analysis, disagreements related to theme interpretations were resolved through discussion among the researchers, and the coding scheme was iteratively refined until the team reached a consensual thematic structure.

\section{Results - Disagreement Patterns}

\begin{figure*}[!bph]
  \centering
  \includegraphics[width=\textwidth]{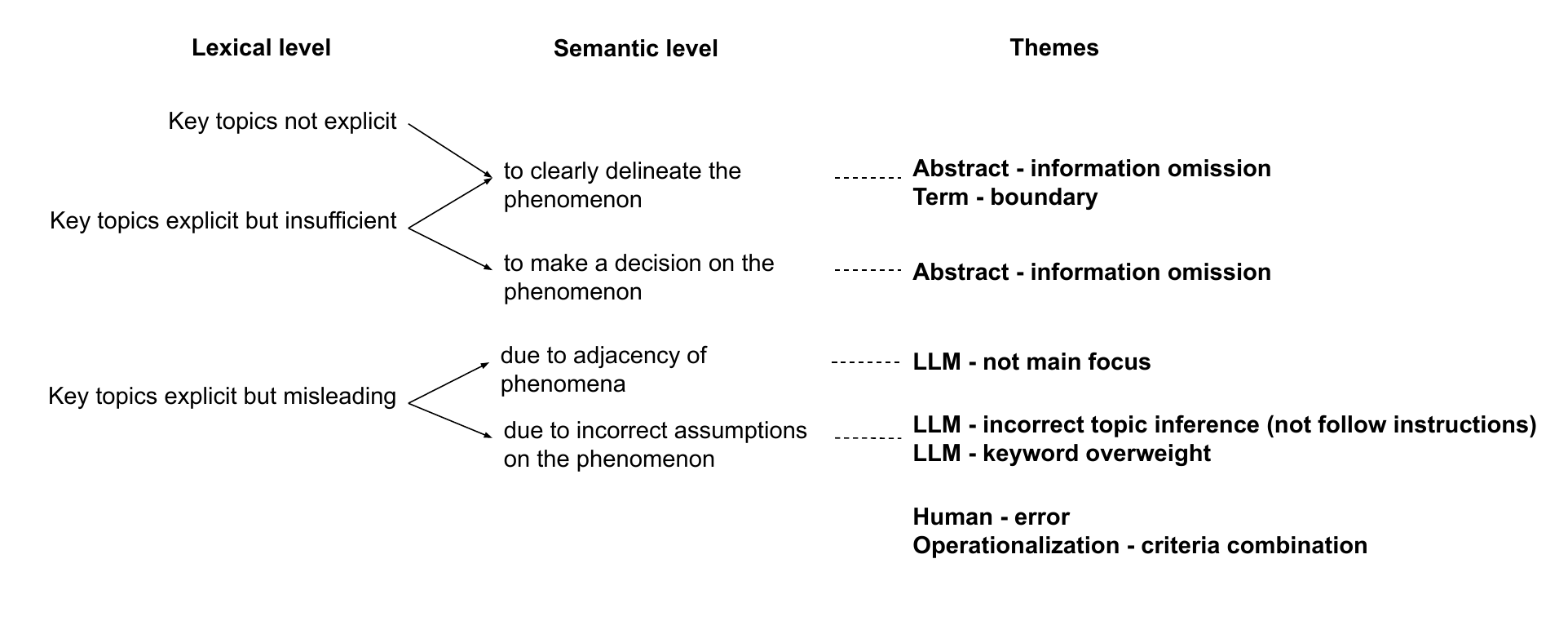}
  \vspace{-10pt}
  \caption{Pathways for themes --The figure shows how lexical cues (e.g., explicit, missing, or misleading topic-related terms) in titles and abstracts (at the \textbf{lexical level}) give rise to different semantic interpretation challenges with respect to the phenomenon of interest (at the \textbf{semantic level}).}
  \label{fig:paths}
\end{figure*}

To answer RQ1, we present a taxonomy of disagreement patterns, where each pattern identifies the location and nature of the issue. The most frequent sources of divergence involved missing explicit terminology in abstracts, conceptual boundary ambiguity, and models over-relying on surface lexical cues. In the rest of this section, we provide illustrative examples for each pattern, classified under two error types: Evidence Loss (False Negative) and False Positive. We operationalize the two error types relative to the human consensus decision.  
\textit{Evidence Loss (EL)} occurs when the LLM excludes a study that humans included, representing a potential loss of relevant evidence. 
\textit{False Positive (FP)} occurs when the LLM includes a study that humans excluded, representing the introduction of potentially irrelevant evidence into the screening set.

When deriving the disagreement patterns themes, we observed that some of them had problems at the lexical level (in the form of the word), or at the semantic level (related to the meaning of the words)~\cite{jm3}. They are illustrated in Figure \ref{fig:paths}. The figure depicts the different pathways from lexical issues to semantic causes or consequences associated with each theme. It does not connect the theme "Human -- error" to the lexical--semantic pathways, as it pertains to human judgment rather than to LLM processing. Similarly, "Operationalization -- criteria combination" pertains to combining screening criteria rules and is not related to lexical--semantic pathways, so it is left disconnected in the figure.

In cases under ``Abstract -- information omission'' and ``Term -- boundary,'' the core issue was at the lexical level: the key topics of the SR were not explicit. As a result, the phenomenon of interest was not clearly delineated. In some situations, the key topics were explicit but insufficient, generating eligibility uncertainty at the semantic level. This applies to some cases under the ``Abstract -- information omission'' pattern. For other patterns, the key topics were explicit but potentially misleading at the semantic level. In some cases, this occurred due to adjacency between the phenomenon of interest and related phenomena, resulting in ``LLM -- not main focus'' pattern. For example, a paper may tangentially address the topic of interest but not as a primary contribution. In other cases, the misleading interpretation stemmed from incorrect assumptions made by the LLM, as represented by the patterns ``LLM -- incorrect topic inference (not follow instructions)'' and ``LLM -- keyword overweight.''

Table~\ref{tab:pattern-distribution} summarizes the occurrence of disagreements per study across the patterns and error types. The most frequent pattern overall was \textit{LLM—Incorrect Topic Inference} (32 instances), followed by \textit{LLM—Not main focus} (23) and \textit{Term—Boundary} (18). Evidence Loss outnumbered False Positives in most patterns, except for \textit{LLM—Not main focus} and \textit{LLM—Incorrect Topic Inference} in certain studies.
In the next sections, we present each pattern individually.

\begin{table}[htbp]
\centering
\caption{Disagreements by pattern, study, and error type. 
EL = Evidence Loss (LLM false negative relative to human consensus); 
FP = False Positive (LLM false positive).}
\label{tab:pattern-distribution}
\scriptsize
\setlength{\tabcolsep}{3.5pt}
\renewcommand{\arraystretch}{1.15}
\resizebox{\textwidth}{!}{%
\begin{tabular}{lrrrrrrrrrrrr}
\toprule
\multirow{2}{*}{\textbf{Pattern}} 
& \multicolumn{2}{c}{\textbf{TP}} 
& \multicolumn{2}{c}{\textbf{ESEE}} 
& \multicolumn{2}{c}{\textbf{WOSS}} 
& \multicolumn{2}{c}{\textbf{GenAIEdu}} 
& \multicolumn{2}{c}{\textbf{AI4T}} 
& \multicolumn{2}{c}{\textbf{AI4TInd}} \\
\cmidrule(lr){2-3}
\cmidrule(lr){4-5}
\cmidrule(lr){6-7}
\cmidrule(lr){8-9}
\cmidrule(lr){10-11}
\cmidrule(lr){12-13}
& EL & FP & EL & FP & EL & FP & EL & FP & EL & FP & EL & FP \\
\midrule
Term--Boundary              
& 3 & -- & -- & 1 & 7 & -- & -- & -- & -- & 2 & 2 & 4 \\

Abstract--Information omission  
& 1 & -- & 3 & -- & 6 & -- & -- & 2 & -- & 1 & -- & -- \\

LLM--Keyword overweight     
& -- & -- & -- & -- & -- & 1 & -- & -- & -- & 1 & -- & 6 \\

LLM--Not main focus         
& -- & 15 & 1 & 1 & 3 & -- & -- & -- & 1 & 2 & -- & -- \\

LLM--Incorrect topic inference   
& 4 & 2 & -- & -- & -- & -- & -- & 13 & 9 & 4 & 1 & 3 \\

Human--Error                
& 1 & -- & -- & -- & 2 & -- & 1 & -- & -- & 1 & -- & -- \\

Operational--Criteria combination 
& 4 & -- & -- & -- & -- & -- & -- & -- & -- & -- & -- & -- \\
\midrule
\textbf{Totals} 
& \textbf{13} & \textbf{17} 
& \textbf{4} & \textbf{2} 
& \textbf{18} & \textbf{1} 
& \textbf{1} & \textbf{15} 
& \textbf{10} & \textbf{11} 
& \textbf{3} & \textbf{13} \\
\bottomrule
\end{tabular}
}%
\end{table}


\subsection{Term - Boundary}
This patter is about disagreement over how broadly or narrowly a concept should be interpreted. This pattern occurs when interpretation of terms fail to delineate construct or topic boundaries, leading to ambiguity in interpreting inclusion and exclusion criteria. LLMs are trained on a wide range of texts that may have broader or narrower interpretations of terms than in the SR. 

\textbf{Evidence Loss}
Evidence loss arises when LLMs interpret key terms more narrowly than intended, excluding papers that human raters would retain.
The WOSS study further highlighted persistent ambiguity around what constitutes participation or contribution in open source contexts. Several papers situated at the margins of these constructs triggered disagreement, with LLMs frequently excluding studies that human reviewers judged as in-scope. 
Similarly, in the TP study, an LLM did not consider "software task effort estimation," "security audits," or "UX design" as software development tasks, whereas human evaluators did, leading to unwarranted exclusions. In AITestIndustry, an LLM did not consider Volvo Trucks to be part of the software industry, whereas for the researcher, any company that develops software is considered part of the software industry.

\textbf{False positive}
Boundary ambiguity can also lead in the opposite direction, where LLMs include papers that fall outside the intended scope. In AITest2023, both LLMs included a paper on vulnerability detection and monitoring, while the human evaluator disagreed. The paper had matched the database search only because it focused on detecting vulnerabilities in test cases, a context that the LLMs failed to account for when assessing relevance. 
In the ESEE study, an LLM classified Use Case Points as an expert-based estimation technique and included a paper, while human raters excluded it because its structured, formula-driven approach to software size calculation does not qualify as expert judgment.

\subsection{Abstract - Information Omission} The title-abstract lacks sufficient information to determine eligibility at the screening stage, leading to divergent uncertainty handling.

\textbf{Evidence Loss}
An example from the ESEE study is a paper on software effort estimation that does not explicitly mention expert judgment in its title or abstract. This led the LLM to exclude the paper, while the human raters included it to avoid losing a potentially relevant study.
In the TP study, the abstract contains a vague sentence regarding time pressure in the section where the results are presented: "The convergence of competition, time pressure, and an engaging environment creates a stimulating atmosphere that significantly drives skill development and enhances the overall educational experience." For human raters, it was clear that the paper should be included in the title/abstract screening. However, both LLMs would exclude this paper.

This pattern was particularly evident in the WOSS study. Several papers discussing participation-related phenomena (e.g., disengagement or interaction patterns) were excluded by the LLMs because the abstracts did not explicitly use terms such as ``contribution'' or ``participation.'' Human raters, however, retained these papers based on contextual interpretation of the study focus.

\textbf{False Positive}
False positives emerged when abstracts lacked sufficient detail to confirm eligibility but contained relevant surface signals. In the WOSS study, the abstracts had subtle cues indicating that the papers were doctoral symposium publications (an exclusion criterion). Human reviewers detected these signals and excluded the papers accordingly, whereas the LLMs included them. These cases indicate that subtle publication-type signals in abstracts may be inconsistently captured by LLMs during screening.

\subsection{LLM - Keyword Overweight} The LLM overweights keyword presence in the abstract.

\textbf{Evidence loss} 
We did not observe clear cases of evidence loss primarily attributable to keyword overweight in our dataset.

\textbf{False Positive}
A particularly clear example of this category occurred in the WOSS study, where an LLM included a paper on participation in a Type 1 diabetes management case study solely because it mentioned open-source software used for insulin delivery. Despite matching keywords, the study was clearly outside the review scope. 
In AITest2023, a paper was included despite it being evident that the study focused on hardware testing, which was explicitly listed in the exclusion criteria. These cases illustrate how strong lexical overlap can override contextual relevance in LLM decision-making.

\subsection{LLM - Not main focus} 
The title/abstract focuses on a topic conceptually adjacent to the topic of interest, rather than directly addressing it.

\textbf{Evidence loss}
In AITest2023, a paper considered the use of LLMs for testing, along with code generation and bug fixing. Since the study was not solely focused on testing, Gemini excluded it, whereas human evaluators decided to retain it for the full-text review phase.
In the WOSS study, several papers centered on themes such as equity, inclusivity, or interaction patterns in OSS communities were excluded by the LLMs because ``participation'' or ``gender'' was not the explicit focal construct. Human reviewers, however, retained these studies for full-text inspection. This suggests that LLMs apply a stricter notion of topical centrality than human screeners.

\textbf{False Positive}
An example from the ESEE study involves a paper addressing a strongly human-centered topic, such as negotiation and the defense of estimates, which could apply to any type of estimation method or technique. The LLMs appear to have been influenced by this focus and decided to include the paper, whereas the human raters excluded it.

In the TP study, GPT-4 identified multiple papers in which time pressure was cited as motivation in the abstract. Still, the abstract failed to establish that the paper contained any evidence of time pressure. Consequently, Gemini correctly excluded these papers.


\subsection{LLM - Incorrect Topic Inference}
This pattern occurs when an LLM infers a topic or contextual framing that is not supported by the paper's title or abstract, leading to decisions that contradict the stated criteria.

\textbf{Evidence Loss}
Evidence loss in this pattern is a result of LLMs' mishandling of negations in exclusion criteria or misjudging the level of specificity required. In the TP study, an LLM frequently provided a correct rationale but still reached the wrong decision, likely due to difficulty handling negation in the exclusion criteria: "The task studied is not a software development task." Although this issue recurred multiple times, Gemini was inconsistent and more often made the correct decision. GPT-4.1-mini did not show this problem. In AITest2023, Gemini excluded a paper on LLM-based testing because it operated at a higher level of abstraction rather than addressing specific methodologies or implementations, a judgment human evaluators did not share. In another case, Gemini excluded a paper on LLM-based test generation for programming benchmarks, incorrectly interpreting it as a study on testing AI rather than on AI-based test generation.

\textbf{False Positive}
False positives in this pattern arise when LLMs infer a plausible but unsupported contextual framing. In AITest2023, both LLMs included several papers focused on "testing of AI" when the objective was to identify papers on "AI for testing." Although the criteria explicitly distinguished between the two directions, the LLMs failed to consistently apply this distinction. A similar issue arose in the GenAIEdu study, where one LLM repeatedly inferred an educational context even when the title and abstract provided no explicit indication of one, leading to incorrect inclusion decisions.

\subsection{Human - Error}	The human revised the screening decision after exposure to the LLM's rationale, which highlighted information that had previously been overlooked. 

\textbf{Evidence loss}
In the TP study, humans overlooked a study that investigated time pressure and improvisation as its main focus in the context of a software team. Both human evaluators missed this study, whereas both LLMs correctly identified it as relevant for inclusion. A similar situation occurred in the GenAIEdu study. In these cases, the LLMs mitigated evidence loss by identifying relevant studies that had been missed during human screening.
A similar borderline case was observed in the WOSS study, where human reviewers initially excluded a paper due to limited explicit framing of participation in the abstract. The LLM, however, flagged the study for inclusion based on contextual cues. Upon closer inspection, the paper was deemed relevant, 
suggesting that model rationales can occasionally surface evidence overlooked during manual screening.

\textbf{False Positive}
We observed one borderline case in AI4T where the human decision may have been influenced by an oversight rather than a judgment call. However, the evidence was not conclusive enough to classify it as human error.

\subsection{Operationalization - Criteria combination} Typically, SRs require that all inclusion criteria are met and none of the exclusion criteria. However, SRs may apply different logic to the criteria. 

\textbf{Evidence Loss}
This occurred in the TP study, where meeting either inclusion criterion was sufficient for inclusion. Occasionally, LLMs would exclude relevant studies based on the fact that only one inclusion criteria was satisfied. Yet, when we conducted further analysis, we noticed that although LLMs default to an AND logic for distinct criteria, they may inadvertently transition to OR logic. This observation is reflected in our recommendations. Upon identifying our incorrect operationalization, the TP study was re-run using Boolean logic, where the LLM evaluated each criterion separately and Boolean logic was applied programmatically, which revealed 4 cases of evidence loss. 

\textbf{False Positive} We did not observe clear cases of false positives attributable to Operationalization in our dataset.


\section{Discussion}

\subsection{Recommendations}
In this subsection, we answer RQ2 by providing recommendations based on the experiences from analyzing disagreements between human raters and LLMs. The recommendations below were derived from the recurring disagreement patterns identified in RQ1. Each recommendation targets one or more observed sources of disagreement between human reviewers and LLMs, as summarized in Table~\ref{tab:pattern-recs}. We have placed these recommendations in the Discussion section, as their impacts require validation in future studies.

\begin{table}[htb]
\centering
\caption{Mapping of disagreement patterns to recommendations.}
\label{tab:pattern-recs}
\footnotesize
\begin{tabular}{lrrrrr}
\toprule
\textbf{Pattern} & \textbf{R1} & \textbf{R2} & \textbf{R3} & \textbf{R4} & \textbf{R5} \\
\midrule
Term--Boundary                    &   &   & \checkmark & \checkmark &   \\
Abstract--Information omission    & \checkmark &   &   & \checkmark & \checkmark \\
LLM--Keyword overweight           & \checkmark & \checkmark & \checkmark &   &   \\
LLM--Not main focus               & \checkmark &   & \checkmark & \checkmark &   \\
LLM--Incorrect topic inference    & \checkmark & \checkmark & \checkmark & \checkmark &   \\
Human--Error                      &   &   &   & \checkmark & \checkmark \\
Operationalization--Criteria comb.&   & \checkmark & \checkmark &   &   \\
\bottomrule
\end{tabular}
\end{table}

\textbf{R1-Run Multiple LLMs}. Prior quantitative work suggests that different LLMs, above a certain size, have similar aggregate screening accuracy \cite{huotalaSESREvalDatasetEvaluating2025}. However, this does not mean individual decision would be the same. We find that running multiple LLMs results in greater diversity. Each model may have its own interpretations and misunderstandings affecting the SR study. Using multiple models helps uncover these misunderstandings and interpretations. However, even after executing multiple models, some misunderstandings may persist across models. In such cases, researchers can consider general misunderstandings that models may have learned from training data as misunderstandings from public forums can enter model weights through training data. 
For instance, a widespread misconception that code reviews primarily catch functional defects, whereas evidence shows they are predominantly maintainability issues, could affect screening decisions across multiple models simultaneously~\cite{siy2001does,mantyla2008types,beller2014modern}. More details on such misconceptions can be found in work by Rainer \cite{rainer2017using}.

\textbf{R2-Run Each Criteria Separate}. 
The general wisdom is that LLMs perform better when the task is smaller. This should also be followed in this context. We have found cases where LLMs do not produce internally consistent results, as the overall decision diverges from individual criteria decisions. Additionally, SRs can include different boolean logic, e.g., rules that are interpreted using OR-logic rather than AND-logic (e.g., for TP-studies). Therefore, it is preferable to feed each criterion separately to the LLM. The combination and final decision should be done via traditional programming using Boolean logic.

\textbf{R3-Define Criteria Unambiguously}. Many disagreements stemmed from underspecified inclusion/exclusion criteria, especially around construct boundaries and domain concepts (e.g., what counts as ``participation,'' ``software testing,'' or ``industry''). Before running LLM-assisted screening, researchers should make the criteria definition complete by explicitly stating (i) key term definitions, (ii) boundary conditions (what is in vs. out), and (iii) common confounders and near-miss cases. This reduces semantic drift and helps LLMs apply the intended scope consistently. 

\textbf{R4-Validate From Boundaries}: 
In SR screening, many studies are straightforward ``includes'' or ``excludes'' and provide limited insight into model behavior. Greater value comes from examining borderline cases near the inclusion boundary, as these are more likely to reveal systematic weaknesses in both human and LLM decision-making.
There is little to learn regarding screening accuracy from these easy decisions. However, the studies that lie on the boundaries of the include/exclude decision are the ones that require attention and are the key to sharpening the decision rules both for humans and 
LLMs. Below are two techniques for identifying these boundaries:

\textbf{R4.1-LLM Disagreement} One can run multiple LLMs and sample all studies where the LLMs disagree with each other. These cases are guaranteed to contain LLM misclassifications. Studying these cases and the reasons provided by the LLMs for their decisions can offer insights into how to improve the process. For example, we observed that some models claimed vulnerability detection is part of software testing, and that a study conducted at Volvo Trucks is not a study done "in the software industry" because, according to the model, Volvo Trucks is not considered part of the software industry. 
We do disagree with those statements; however, these viewpoints can substantially reduce the number of studies that need to be re-examined manually, allowing human reviewers to focus their attention on a smaller subset of borderline cases. Additionally, such insights on disagreements help clarify the criteria in R3-Define Criteria Unambiguously.

\textbf{R4.2-LLM Probability} One can also ask the LLMs to output probabilities instead of a binary decision for each criterion. This forces LLMs to reveal any uncertainty they may have. Although we have noticed that it is quite rare to get an LLM to output a 0.5 probability, there are a fair number of cases where probabilities of 0.7 and 0.3 are output. When using probabilities, one should start from the middle, that is, the 0.5 range, and move both up and down to identify borderline cases. This can also help determine how the probability threshold should be set for the decision. One can also average probabilities from multiple LLMs and attempt to identify a "shoulder" for the cutoff points. A shoulder is the point in the density graph where the probability quickly drops.

\textbf{R5-Double Check Human Decisions}
Beyond serving as an automated classifier, the LLM can be used as an external opinion mechanism to stress-test human screening decisions and identify studies that warrant careful re-examination and confirmation of classifications. Rather than replacing human judgment, this external perspective can support reviewers, particularly novice researchers or teams with limited human resources, by highlighting borderline cases that deserve closer scrutiny. In this study, we identified human errors that could have resulted in evidence loss had the LLMs not flagged certain studies.

On the face value, the above recommendations make sense to all study authors who have experience in executing multiple SRs and who have all recently been active in investigating the use of large language models (LLMs) in SRs \cite{huotalaPromiseChallengesUsing2024a,huotalaSESREvalDatasetEvaluating2025,AISysRev,felizardo2025difficulties}.
Yet, the impact in terms of empirical evidence of the recommendations is currently unclear, as each of them would require a quantitative study of its own. 
Thus, we welcome the community as a whole to continue investigating the impact of the recommendations.


\subsection{Comparison to related work}
As pointed out many works have quantitatively evaluated the screening accuracy of LLMs \cite{huotalaPromiseChallengesUsing2024a,felizardoChatGPTApplicationSystematic2024,huotalaSESREvalDatasetEvaluating2025,petersen2025road,thode2025exploring,lieberum2025large,kim2025evaluating,sandner2025evaluating,chan2025comprehensive} but far fewer studies have qualitative analysed the problems in LLM screening or provided mitigation strategies.  

Recently, Madeyski et al. \cite{madeyski2025llm4screenlit} outlined recommendations on reporting the quantitative results of large language model (LLM) use in title-abstract screening. For example, in addition to precision and recall, one should report the Matthews correlation coefficient. We on the other hand provide recommendations on LLM usage rather than result reporting. 
Felizardo \cite{felizardo2025difficulties} reports on the difficulties in using LLMs in SRs based on two studies. While their scope is wider than that of this paper, there is a similar finding regarding LLM prompt sensitivity. In our work, we provided more details on this aspect through an error-class analysis based on six studies and also offered a recommendation (R3) to address it.
Petersen \cite{petersen2024case} tackled the problem of misusing the term "case study" by original study authors and employed an LLM to detect such misclassifications. This aligns with our recommendation R5 and provides a more detailed context on applying LLMs to double-check human work.
Thode \cite{thode2025exploring} conducted primarily quantitative work on two SRs but also analyzed LLM mistakes. They reached some similar findings to ours: using unmodified and unimproved screening criteria (R3) may be problematic, and employing multiple LLMs—though artificially increasing recall—can help avoid evidence loss. 

To summarize, no existing software engineering study focuses exclusively on qualitative analysis of LLM errors in SR screening as we have done in this paper. However, similar insights were found as sub-analysis from the prior works.


\subsection{Limitations}

This study has limitations that should be considered when interpreting the findings. First, our analysis is limited to the title-abstract screening stage of SRs. Disagreement patterns at other stages, such as full-text screening or data extraction, may differ and warrant separate investigation. 
Second, while four studies employed multiple independent researchers, the two ongoing studies relied on a single rater, which may introduce bias in the coding of disagreements. This trade-off was deliberate, given the larger number of papers in those studies. 
Third, all LLM screening was conducted in zero-shot mode using specific models. Different prompting strategies, such as few-shot or chain-of-thought prompting, may yield different disagreement patterns and are not reflected in our taxonomy. Furthermore, the results reflect the specific model versions available at the time of the study and may evolve as LLM capabilities continue to improve.
Fourth, the taxonomy and recommendations were derived inductively from a limited set of studies. While the involvement of six researchers in the analysis process strengthens credibility, the patterns identified should be considered a starting point rather than an exhaustive account of LLM disagreement modes in title-abstract screening.
Fifth, our findings are grounded in software engineering SRs, which may limit direct generalization to other domains where terminology, inclusion criteria, and topic boundaries differ.
Additionally, three of the six subject studies (GenAIEdu, AI4T, AI4TInd) were ongoing or unpublished at the time of analysis and had not undergone full peer review. While this introduces some uncertainty about the stability of their inclusion criteria, it also reflects realistic conditions under which LLM-assisted screening is likely to be used in practice.
Finally, the use of human consensus as the reference decision may raise concerns. Human screening itself is subject to error and interpretation variability. Therefore, some disagreements attributed to LLMs may reflect ambiguity in the primary studies or limitations in the human judgment process.

\section{Conclusion}
LLMs have the potential to support researchers in title-abstract screening, which is a critical step in SRs. In this paper, we investigate how LLMs fail during screening. Through qualitative analysis of disagreements between LLMs and human raters across six software engineering SRs, we derived a taxonomy of seven disagreement patterns. Based on these patterns, we proposed five recommendations to help design more reliable LLM-assisted screening workflows.

Our findings suggest that LLM failures result from recurring, identifiable causes, such as boundary ambiguity in key terms, information omission in abstracts, keyword overemphasization, and incorrect topic inference. Recognizing these patterns allows researchers to take targeted preventive measures, such as validating semantic understanding before deployment, running multiple LLMs, evaluating each criterion separately, and focusing validation efforts on borderline cases. We expect this work to contribute to more informed and transparent use of LLMs in evidence synthesis and to encourage future research to build on this taxonomy using larger, more diverse sets of SRs.

\subsubsection{Data Availability.} \url{https://dx.doi.org/10.5281/zenodo.20695824} 

\subsubsection{\ackname}
This work was supported by the Research Council of Finland (Grant id: 359861) and Strategic Research Council of Research Council of Finland (Grant ID 358471). This study was financed in part by the Coordenação de Aperfeiçoamento de Pessoal de Nível Superior - Brasil (CAPES-PROEX) - Finance Code 001. This work was also partially supported by CNPq (processes 314797/2023-8, 443934/2023-1, and 445029/2024-2), Amazonas State Research Support Foundation - FAPEAM - through the POSGRAD project 2025/2026 and the Federal University of Mato Grosso do Sul (UFMS).

\subsubsection{\discintname}
The authors have no competing interests to declare that are relevant to the content of this article.


\bibliographystyle{templates/springer/splncs04}
\bibliography{9_refs}


\end{document}